\documentstyle[preprint,aps,graphicx,floats]{revtex}
\def\ba{\begin{eqnarray}}
\def\ea{\end{eqnarray}}
\def\bq{\begin{equation}}
\def\eq{\end{equation}}
\def\lsim{\mathrel{\raisebox{-.6ex}{$\stackrel{\textstyle<}{\sim}$}}}
\def\gsim{\mathrel{\raisebox{-.6ex}{$\stackrel{\textstyle>}{\sim}$}}}

\begin{document}

\thispagestyle{empty}

\newcommand{\sla}[1]{/\!\!\!#1}

\preprint{
\font\fortssbx=cmssbx10 scaled \magstep2
\hbox to \hsize{
\hskip.5in \raise.1in\hbox{\fortssbx University of Wisconsin - Madison}
\hfill\vtop{\hbox{\bf MADPH-00-1157}
            \hbox{CERN-TH/2000-039} 
            \hbox{February 2000}
            \hbox{ } } 
}
}

\vskip.5in

\title{ Measuring Higgs boson couplings at the LHC } 

\author{
D.~Zeppenfeld$^1$, R.~Kinnunen$^2$, A.~Nikitenko$^3$, E.~Richter-W\c{a}s$^4
$
} 

\address{ 
$^1$Department of Physics, University of Wisconsin, Madison, WI 53706, USA 
and\\
CERN, TH Division, 1211 Geneva 23, Switzerland \\
$^2$Helsinki Institute of Physics, Helsinki, Finland\\
$^3$Helsinki Institute of Physics, Helsinki, Finland, on leave from ITEP, 
Moscow, Russia\\
$^4$CERN, IT Division, 1211 Geneva 23, Switzerland and \\
Institute of Computer Science, Jagellonian University, Institute of 
Nuclear Physics,\\
  30-055 Krakow, ul. Kawiory 26a, Poland
}  

\maketitle 

\begin{abstract}
For an intermediate mass Higgs boson with SM-like couplings the LHC
allows observation of a variety of decay channels in production by gluon 
fusion and weak boson fusion. Cross section ratios provide measurements 
of various ratios of Higgs couplings, with accuracies of order 15\%
for 100~fb$^{-1}$ of data in each of the two LHC experiments.
For Higgs masses above 120~GeV, minimal assumptions on the Higgs sector 
allow for an indirect measurement of the total Higgs boson width with an 
accuracy of 10 to 20\%, and of the $H\to WW$ partial width with an accuracy
of about 10\%.
\end{abstract} 


\section{Introduction}

Investigation of the symmetry breaking mechanism of the electroweak 
$SU(2)\times U(1)$ gauge symmetry will be one of the prime tasks of the
LHC. Correspondingly, major  efforts have been concentrated on devising 
methods for Higgs boson discovery, for the entire mass range 
allowed within the Standard Model (SM) (100~GeV$\lsim m_H \lsim 1$~TeV,
after LEP2), and for Higgs boson search in extensions of the SM,
like the MSSM~\cite{ATLAS,CMS}. While observation of one or more Higgs 
scalar(s) at the LHC appears assured, discovery will be followed by
a more demanding task: the systematic investigation of Higgs boson 
properties. Beyond observation of the various CP even and CP odd scalars 
which nature may have in store for us, this means the determination of 
the couplings of the Higgs boson to the known fermions and gauge bosons,
i.e. the measurement of $Htt$, $Hbb$, $H\tau\tau$ and $HWW$, $HZZ$, 
$H\gamma\gamma$ couplings, to the extent possible. 

Clearly this task very much depends on the expected Higgs boson mass. For
$m_H>200$~GeV and within the SM, only the $H\to ZZ$ and $H\to WW$ channels 
are expected to be observable, and the two gauge boson modes are related 
by SU(2). Above $m_H\approx 250$~GeV, where detector effects will no longer
dominate the mass resolution of the $H\to ZZ\to 4\ell$ resonance,
additional information is expected from a direct measurement of the total 
Higgs boson width, $\Gamma_H$. A much richer spectrum of decay modes
is predicted for the intermediate mass range, i.e. if a SM-like
Higgs boson has a mass between the reach of LEP2 ($\lsim 110$~GeV) and 
the $Z$-pair threshold. The main reasons for focusing on this range are 
present indications from electroweak precision
data, which favor $m_H<250$~GeV~\cite{EWfits}, as well as expectations
within the MSSM, which predicts the lightest Higgs boson to have a mass
$m_h\lsim 130$~GeV~\cite{mhMSSM}.

Until recently, the prospects of detailed and model independent 
coupling measurements at the LHC were considered somewhat 
remote~\cite{snowmass_H}, because few promising search channels were
known to be accessible, for any given Higgs boson mass. 
Taking ATLAS search scenarios as an example, these were~\cite{ATLAS}
\ba 
\label{eq:ggHAA}
gg\to H\to \gamma\gamma\;, \qquad 
{\rm for}\quad m_H \lsim 150~{\rm GeV}\;, \\
\label{eq:ggHZZ}
gg\to H\to ZZ^*\to 4\ell\;,\qquad 
{\rm for}\quad m_H \gsim 130~{\rm GeV}\;, 
\ea
and
\bq
\label{eq:ggHWW}
gg\to H\to WW^*\to \ell\bar\nu\bar\ell\nu\;, \qquad 
{\rm for}\quad m_H \gsim 150~{\rm GeV}\;, 
\eq
with the possibility of obtaining some additional information from processes 
like $WH$ and/or $t\bar tH$ associated production with subsequent 
$H\to\bar bb$ and $H\to\gamma\gamma$ decay for Higgs boson masses near 
100~GeV. Throughout this paper, ``$gg\to H$'' stands for inclusive 
Higgs production, which is
dominated by the gluon fusion process for a SM-like Higgs boson.

This relatively pessimistic outlook is changing considerably now, due to the
demonstration that weak boson fusion is a promising Higgs production channel
also in the intermediate mass range. Previously, this channel 
had only been explored for Higgs masses above 300~GeV. 
Specifically, it was recently shown in parton 
level analyses that the weak boson fusion channels, with subsequent Higgs
decay into photon pairs~\cite{RZ_gamgam,R_thesis},
\bq
\label{eq:wbfHAA}
qq\to qqH,\;H\to \gamma\gamma\;, \qquad 
{\rm for}\quad m_H \lsim 150~{\rm GeV}\;,
\eq
into $\tau^+\tau^-$ pairs~\cite{R_thesis,RZ_tautau_lh,RZ_tautau_ll},
\bq
\label{eq:wbfHtautau}
qq\to qqH,\;H\to \tau\tau\;, \qquad 
{\rm for}\quad m_H \lsim 140~{\rm GeV}\;,
\eq
or into $W$ pairs~\cite{R_thesis,RZ_WW}
\bq
\label{eq:wbfHWW}
qq\to qqH,\;H\to WW^{(*)}\to e^\pm \mu^\mp /\!\!\!{p}_T\;, 
\qquad {\rm for}\quad m_H \gsim 120~{\rm GeV}\;,
\eq
can be isolated at the LHC. Preliminary analyses, which try to extend
these parton level results to full detector simulations, look 
promising~\cite{sasha}. The weak boson fusion channels utilize the 
significant background reductions which are expected from 
double forward jet tagging~\cite{Cahn,BCHP,DGOV} and central jet vetoing 
techniques~\cite{bjgap,bpz_mj}, and promise low
background environments in which Higgs decays can be studied in detail.  
The parton level results predict highly significant signals with 
(substantially) less than 100~fb$^{-1}$.

The prospect of observing several Higgs production and decay 
channels, over the entire intermediate mass range, suggests a reanalysis of 
coupling determinations at the LHC~\cite{snowmass_H}. In this paper we 
attempt a first such 
analysis, for the case where the branching fractions of an intermediate
mass Higgs resonance are fairly similar to the SM case, i.e. we analyze
a SM-like Higgs boson only. We make use of the previously published analyses
for the inclusive Higgs production channels~\cite{ATLAS,CMS} and of the 
weak boson fusion 
channels~\cite{RZ_gamgam,R_thesis,RZ_tautau_lh,RZ_tautau_ll,RZ_WW}.
The former were obtained by the experimental collaborations and include 
detailed detector simulations. The latter are based on parton level results,
which employ full QCD tree level matrix elements for all signal and 
background processes. We will not discuss here differences in the 
performance expected for the ATLAS and CMS detectors nor details in the 
theoretical assumptions which lead to different estimates for expected 
signal and background rates. The reader is referred to the original 
publications from which numbers are extracted.
In Section~\ref{sec2} we summarize expectations for the various 
channels, including expected accuracies for cross section measurement of
the various signals for an integrated luminosity of 100~fb$^{-1}$. 
Implications for the determination of coupling ratios and the measurement 
of Higgs boson (partial) decay widths are then obtained in
Section~\ref{sec3}. A final summary is given in Section~\ref{sec4}.

\section{Survey of intermediate mass Higgs channels}
\label{sec2}

\begin{table*}[thb]
\caption{Number of expected events for the inclusive SM $H\to \gamma\gamma$ 
signal and expected backgrounds, assuming an integrated luminosity of 
100~${\rm fb}^{-1}$ and high luminosity performance. Numbers correspond 
to optimal $\gamma\gamma$ invariant mass windows for CMS and ATLAS.
The expected relative statistical errors on the signal cross section 
are given for the individual experiments and are combined in the last line.
}
\vspace{0.15in}
\label{table:gg.AA}
\begin{tabular}{c|c|cccccc}
    & $m_H$ & 100  & 110  & 120  & 130  & 140  & 150  \\
\hline
CMS~\protect\cite{CMS_ecal_tdr,katri_lassila} 
    & $N_S$ &   865 &  1038 &  1046 &   986 &  816 &  557 \\
    & $N_B$ & 29120 & 22260 & 16690 & 12410 & 9430 & 7790 \\
    & $\Delta\sigma_H/\sigma_H$ 
            & 20.0\% & 14.7\% & 12.7\% & 11.7\% & 12.4\% & 16.4\% \\
\hline
ATLAS~\protect\cite{ATLAS} 
      & $N_S$ &  1045 &  1207 &  1283 &  1186 &   973 &   652 \\
      & $N_B$ & 56450 & 47300 & 39400 & 33700 & 28250 & 23350 \\
    & $\Delta\sigma_H/\sigma_H$ 
              & 22.9\% & 18.2\% &15.7\% & 15.7\% & 17.6\% & 23.8\% \\
\hline 
Combined & $\Delta\sigma_H/\sigma_H$ 
              & 15.1\% & 11.4\% & 9.9\% & 9.4\% & 10.1\% & 13.5\% \\
\end{tabular}
\end{table*}

The various Higgs channels listed in Eqs.~(\ref{eq:ggHAA}--\ref{eq:wbfHWW})
and their observability at the LHC have all been discussed in the 
literature. Where available, we give values as presently quoted by the 
experimental collaborations. In order to compare
the accuracy with which the cross sections of different Higgs production
and decay channels can be measured, we need to unify these results. For
example, $K$-factors of unity are assumed throughout. Our goal 
in this section is to obtain reasonable estimates for the relative errors,
$\Delta\sigma_H/\sigma_H$, which are expected after collecting 
100~${\rm fb}^{-1}$ in each the ATLAS and the CMS detector, i.e. we estimate 
results after a total of 200~${\rm fb}^{-1}$ of data have been collected 
at the LHC. Presumably these data will be taken with a mix of both low 
and high luminosity running. 
 
We find that the measurements are largely dominated by statistical errors.
For all channels, event rates with 200~${\rm fb}^{-1}$ of data will be large 
enough to use the Gaussian approximation for statistical errors. The 
experiments measure the signal cross section by separately determining the 
combined signal + background rate, $N_{S+B}$, and the expected number of 
background events, $\langle N_B\rangle$. The signal cross section is then 
given by
\bq
\sigma_H = {N_{S+B}-\langle N_B\rangle \over \epsilon\int{\cal L}dt }=
{N_S\over \epsilon\int{\cal L}dt }\;,
\eq
where $\epsilon$ denotes efficiency factors.
Thus the statistical error is given by
\bq
{\Delta\sigma_H\over\sigma_H} = {\sqrt{N_{S+B}}\over N_S} = 
{\sqrt{N_S+N_B}\over N_S}\;,
\eq
where in the last step we have dropped the distinction between the 
expected and the actual number of background events.
Systematic errors on the background rate are added in quadrature to the
background statistical error, $\sqrt{N_B}$, where appropriate.

Well below the $H\to WW$ threshold, the search for $H\to\gamma\gamma$ events
is arguably the cleanest channel for Higgs discovery. LHC detectors have 
been
designed for excellent two-photon invariant mass resolution, with this 
Higgs signal in mind. We directly take the expected signal and background
rates for the inclusive  $H\to\gamma\gamma$ search from the detailed studies
of the CMS and ATLAS collaborations~\cite{CMS_ecal_tdr,katri_lassila,ATLAS}, 
which were performed 
for an integrated luminosity of 100~${\rm fb}^{-1}$ in each detector. 
Expectations are summarized in Table~\ref{table:gg.AA}. Rates correspond to 
not including a $K$-factor for the expected signal and background cross 
sections in CMS and ATLAS. 
Cross sections have been determined with MRS (R1) parton
distribution functions (pdf's) for CMS, while ATLAS numbers are based on 
CTEQ2L pdf's. 

The inclusive $H\to\gamma\gamma$ signal will be observed as a narrow
$\gamma\gamma$ invariant mass peak on top of a smooth background 
distribution.
This means that the background can be directly measured from the very high
statistics background distribution in the sidebands. We expect any 
systematic errors on the extraction of the signal event rate to be 
negligible compared to the statistical 
errors which are given in the last row of Table~\ref{table:gg.AA}.
With  100~${\rm fb}^{-1}$ of data per experiment 
$\sigma(gg\to H)\cdot B(H\to\gamma\gamma)$ can be determined with a relative
error of 10 to 15\% for Higgs masses between 100 and 150~GeV. Here we 
do not include additional systematic errors, e.g. from the luminosity
uncertainty or from higher order QCD corrections, because we will mainly 
consider cross section ratios in the final analysis in the next Section. 
These systematic errors largely cancel in the cross section ratios. 
Systematic errors common to several channels will be considered later, where
appropriate.

A Higgs search channel with a much better signal to background ratio, 
at the price of lower statistics, however, is available 
via the inclusive search for $H\to ZZ^*\to 4\ell$ events. Expected event
numbers for 100~${\rm fb}^{-1}$ in both ATLAS~\cite{ATLAS} and 
CMS~\cite{CMS_HZZ} are listed in 
Table~\ref{table:gg.ZZ}. These numbers were derived using CTEQ2L pdf's 
and are corrected to contain no QCD K-factor. 
For those Higgs masses where no ATLAS or CMS prediction is available, we
interpolate/extrapolate the results for the nearest Higgs mass, taking 
the expected $H\to ZZ^*$ branching ratios into account for the signal.
Similar to the case of
$H\to \gamma\gamma$ events, the signal is seen as a narrow peak in the 
four-lepton invariant mass distribution, i.e. the background can be 
extracted directly from the signal sidebands.
The combined relative error on the measurement
of $\sigma(gg\to H)\cdot B(H\to ZZ^*)$ is listed in the last line of 
Table~\ref{table:gg.ZZ}. For Higgs masses in the 130--150~GeV range, and
above $Z$-pair threshold, a 10\% statistical error on the 
cross section measurement is possible.
In the intermediate range, where $H\to WW$ dominates, and for lower Higgs 
masses, where the Higgs is expected to dominantly decay into $\bar bb$,
the error increases substantially.

\begin{table*}[t]
\caption{Number of expected events for the inclusive SM 
$H\to ZZ^*\to \ell^+\ell^-\ell^+\ell^-$ 
signal and expected backgrounds, assuming an integrated luminosity of 
100~${\rm fb}^{-1}$ and high luminosity performance. Numbers correspond to 
optimal four-lepton invariant mass windows for CMS and ATLAS and to the 
combined total. Rates in parentheses correspond to numbers interpolated,
according to $H\to ZZ^*$ branching ratios for the signal. 
The expected relative statistical errors on the signal cross section are 
given for each experiment and are combined in the last line.}
\vspace{0.15in}
\label{table:gg.ZZ}
\begin{tabular}{cc|ccccccc}
    & $m_H$   &   120  &    130  & 140 &  150  & 160 & 170   & 180  \\
\hline
CMS~\protect\cite{CMS_HZZ}
      & $N_S$ &  19.2  &   55.3  & (99)& 131.4 & (48)& 29.4  & (76.5)\\
      & $N_B$ &  12.9  &   17.1  & (20)&  22.5 & (26)& 27.5  & (27)  \\
    & $\Delta\sigma_H/\sigma_H$ 
            & 29.5\% & 15.4\% &  11.0\% & 9.4\% & 17.9\% & 25.7\% & 13.3\% \\
\hline
ATLAS~\protect\cite{ATLAS}
      & $N_S$ &  10.3  &   28.7  & (51)& 67.6  & (31)& 19.1  &  49.7 \\
      & $N_B$ &  4.44  &   7.76  & (8) & 8.92  & (8) & 8.87  &  8.81 \\
    & $\Delta\sigma_H/\sigma_H$ 
     & 37.3\% & 21.0\% &  15.1\%& 12.9\% & 20.1\%& 27.7\% & 15.4\% \\
\hline 
Combined 
&$\Delta\sigma_H/\sigma_H$  
            & 23.1\% & 12.4\%  & 8.9\%&  7.6\% & 13.4\% & 18.8\% & 10.1\% \\
\end{tabular}
\end{table*}

\begin{table}[thb]
\vspace{-0.15in}
\caption{Number of expected events for the inclusive SM 
$H\to WW^*\to \ell^+\nu\ell^-\bar\nu$ 
signal and expected backgrounds, assuming an integrated luminosity of 
30~${\rm fb}^{-1}$. Numbers correspond to optimized cuts, varying with the 
mass of the Higgs boson being searched for. The expected relative errors 
on the signal cross section are given for each experiment, 
separating the statistical error, the effect of a systematic 5\% error of the 
background level, and the two added in quadrature. The combined error for the
two experiments assumes 100\% correlation of the systematic errors on the 
background determination.
}
\vspace{0.15in}
\label{table:gg.WW}
\begin{tabular}{cc|cccccccc}
      & $m_H$ & 120  &  130   &   140  &   150  
                                       &   160  &   170  &   180  &   190  
\\
\hline
CMS~\protect\cite{DittDrein}
    & $N_S$ &   44 &  106   &   279  &   330  & 468 & 371 & 545 & \\
    & $N_B$ &  272 &  440   &   825  &   732  & 360 & 360 & 1653 & \\
    & $\Delta\sigma_H/\sigma_H$(stat.)     
 & 40.4\% & 22.0\% & 11.9\% & 9.9\% & 6.1\%& 7.3\%&8.6\%&  \\
    & $\Delta\sigma_H/\sigma_H$(syst.)     
 & 30.9\% & 20.8\% & 14.8\% & 11.1\% & 3.8\%& 4.9\%&15.2\%&  \\
    & $\Delta\sigma_H/\sigma_H$(comb.)     
 & 50.9\% & 30.3\% & 19.0\% & 14.9\% & 7.3\%& 8.8\%&17.4\%& 20.6\% \\
\hline
ATLAS~\protect\cite{ATLAS}
    & $N_S$ & & &  &  240   &   400  &   337  &   276  &   124  \\
    & $N_B$ & & &  &  844   &   656  &   484  &   529  &   301  \\
    & $\Delta\sigma_H/\sigma_H$ (stat.)     
     &  & &  & 13.7\% & 8.1\% & 8.5\% & 10.3\% & 16.6\% \\
    & $\Delta\sigma_H/\sigma_H$ (syst.)     
     &  & &  & 17.6\% & 8.2\% & 7.2\% & 9.6\% & 12.1\% \\
    & $\Delta\sigma_H/\sigma_H$ (comb.)     
 & 50.9\% & 30.3\% & 19.0\% & 22.3\% & 11.5\% & 11.1\% & 14.1\% & 20.6\% \\
\hline 
Combined 
    &$\Delta\sigma_H/\sigma_H$  (comb.) & 
 42.1\% & 26.0\% & 17.0\% & 14.8\% & 7.0\% & 8.0\% & 13.6\% & 16.9\% \\
\end{tabular}
\end{table}

Above $m_H\approx 135$~GeV, $H\to WW^{(*)}$ becomes the dominant SM Higgs
decay channel. The resulting inclusive $WW\to\ell^+\nu\ell^-\bar\nu$ signal
is visible above backgrounds, after exploiting the characteristic
lepton angular correlations for spin zero decay into $W$ pairs near 
threshold~\cite{DittDrein}. The inclusive channel, which is dominated by 
$gg\to H\to WW$, has been analyzed by ATLAS for $m_H\geq 150$~GeV and for
integrated luminosities of 30 and 100~${\rm fb}^{-1}$~\cite{ATLAS}
and by CMS for $m_H\geq 120$~GeV and 
30~${\rm fb}^{-1}$~\cite{DittDrein}. The expected event numbers for 
30~${\rm fb}^{-1}$ are listed in Table~\ref{table:gg.WW}.
The numbers are derived without QCD K-factors and use CTEQ2L for ATLAS and 
MRS(A) pdf's for CMS results. 

Unlike the two previous modes, the two missing neutrinos in the $H\to WW$ 
events do not allow for a reconstruction of the narrow Higgs mass peak. 
Since the Higgs signal is only seen as a broad enhancement of the expected 
background rate in lepton-neutrino transverse mass distributions, with 
similar shapes of signal and background after application of all cuts, a 
precise determination
of the background rate from the data is not possible. Rather one has to rely
on background measurements in phase space regions where the signal is weak,
and extrapolation to the search region using NLO QCD predictions. The 
precise
error on this extrapolation is unknown at present, the assumption of a 5\%
systematic background uncertainty appears optimistic but attainable. 
It turns out that with 30~${\rm fb}^{-1}$ already, the systematic error
starts to dominate, because the background exceeds the signal rate by
factors of up to 5, depending on the Higgs mass. Running at high 
luminosity makes matters worse, because the less efficient reduction of 
$\bar tt$ backgrounds, due to less stringent $b$-jet veto criteria, 
increases the 
background rate further. Because of this problem we only present results
for 30~${\rm fb}^{-1}$ of low luminosity running in Table~\ref{table:gg.WW}.
Since neither of the LHC collaborations has presented predictions for the
entire Higgs mass range, we take CMS simulations below 150~GeV and ATLAS
results at 190~GeV, but divide the resultant statistical errors by a 
factor $\sqrt{2}$, to take account of the presence of two experiments.
Between 150 and 180~GeV we combine both experiments, assuming 100\%
correlation in the systematic 5\% normalization error of the background.

\begin{table*}[htb]
\vspace{0.25in}
\caption{Number of expected $\gamma\gamma jj$ events from the
$qq\to qqH,\; H\to \gamma\gamma$ weak boson fusion
signal and expected backgrounds, assuming an integrated luminosity of 
100~${\rm fb}^{-1}$. Numbers correspond to optimal $\gamma\gamma$ invariant 
mass windows for CMS and ATLAS and to the combined total, as projected
from the parton level analysis of Refs.~\protect\cite{RZ_gamgam,R_thesis}. 
The expected relative statistical errors on the signal cross section are 
given for each experiment and are combined in the last line.}
\vspace{0.15in}
\label{table:WBF.AA}
\begin{tabular}{c|c|cccccc}
    & $m_H$ &       100  & 110  & 120  & 130  & 140  & 150  \\
\hline
projected CMS & $N_S$ &       37   &  48  &  56  &  56  &  48  &  33 \\
performance   & $N_B$ &       33   &  32  &  31  &  30  &  28  &  25 \\
    & $\Delta\sigma_H/\sigma_H$     
            & 22.6\% & 18.6\% & 16.7\% & 16.6\% & 18.2\%& 23.1\% \\
\hline
projected ATLAS & $N_S$ &     42   &  54  &  63  &  63  &  54  &  37 \\
performance     & $N_B$ &     61   &  60  &  56  &  54  &  51  &  46 \\
    & $\Delta\sigma_H/\sigma_H$     
            & 24.2\% & 19.8\% & 17.3\% & 17.2\% & 19.0\% & 24.6\% \\
\hline 
combined 
& $\Delta\sigma_H/\sigma_H$  
            & 16.5\% & 13.6\% & 12.0\% & 11.9\% & 13.1\% & 16.8\% \\
\end{tabular}
\end{table*}

The previous analyses are geared towards measurement of the inclusive 
Higgs production cross section, which is is dominated by the gluon fusion
process. 15 to 20\% of the signal sample, however, is expected to arise
from weak boson fusion, $qq\to qqH$ or corresponding antiquark initiated 
processes. The weak boson fusion component can be isolated 
by making use of the two forward tagging jets which are present in these
events and by vetoing additional central jets, which are unlikely to arise
in the color singlet signal process~\cite{bjgap}. A more detailed discussion
of these processes can be found in Ref.~\cite{R_thesis} from which most of 
the following numbers are taken.

\begin{table*}[thb]
\vspace{0.3in}
\caption{Number of expected signal and background events for the
$qq\to qqH\to\tau\tau jj$ channel, for 100~${\rm fb}^{-1}$ and two
detectors. Cross sections are added for
$\tau\tau\to \ell^\pm h^\mp\sla p_T$ and $\tau\tau \to e^\pm \mu^\mp\sla 
p_T$
events as given in Refs.~\protect\cite{R_thesis,RZ_tautau_ll}. 
The last line gives the expected statistical relative error on the 
$qq\to qqH,\; H\to \tau\tau$ cross section. }
\label{table:WBF.tautau}
\vspace{0.15in}
\begin{tabular}{c|cccccc}
    $m_H$ &       100  & 110  & 120  & 130  & 140  & 150  \\
\hline
         $N_S$ &  211  & 197  & 169  & 128  &  79  & 38   \\
         $N_B$ &  305  & 127  & 51   &  32  &  27  & 24   \\
$\Delta\sigma_H/\sigma_H$ &
                  10.8\% & 9.1\% & 8.8\% & 9.9\% & 13.0\% & 20.7\% \\
\end{tabular}
\end{table*}

The $qq\to qqH,\;H\to\gamma\gamma$ process was first analyzed in 
Ref.~\cite{RZ_gamgam}, where cross sections for signal and background
were obtained with full QCD tree level matrix elements. The parton level
Monte Carlo determines all geometrical acceptance corrections. Additional
detector effects were included by smearing parton and photon 4-momenta with 
expected detector resolutions and by assuming trigger, identification and 
reconstruction efficiencies of 0.86 for each of the two tagging jets and 
0.8 for each photon. Resulting cross sections were presented 
in Ref.~\cite{R_thesis} for a fixed $\gamma\gamma$ invariant mass window of 
total width $\Delta m_{\gamma\gamma}=2$~GeV. We correct these numbers for 
$m_H$ dependent mass resolutions in the experiments. We take $1.4\sigma$ mass
windows, as given in Ref.~\cite{ATLAS} for high luminosity running,
which are expected to contain 79\% of the signal events for ATLAS.
The 2~GeV window for $m_H=100$~GeV at CMS~\cite{CMS_ecal_tdr,katri_lassila} is 
assumed 
to scale up like the ATLAS resolution and assumed to contain 70\% of the 
Higgs signal. The expected total signal and background rates for
100~${\rm fb}^{-1}$ and resulting relative errors for the extraction of the
signal cross section are given in Table~\ref{table:WBF.AA}. Statistical 
errors only are considered for the background subtraction, since the 
background level can be measured 
independently by considering the sidebands to the Higgs boson peak.

The next weak boson fusion channel to be considered is $qq\to qqH,\;
H\to \tau\tau$. Again, this channel has been analyzed at the parton level,
including some estimates of detector effects, as discussed for the 
$H\to \gamma\gamma$ case. Here, a lepton identification efficiency of 0.95
is assumed for each lepton $\ell=e,\;\mu$. 
Two $\tau$-decay modes have been considered so 
far: $H\to\tau\tau \to \ell^\pm h^\mp\sla p_T$~\cite{RZ_tautau_lh}
and $H\to\tau\tau \to e^\pm \mu^\mp\sla p_T$~\cite{RZ_tautau_ll}. These 
analyses were performed for low luminosity running. Some deterioration at 
high luminosity is expected, as in the analogous $H/A\to\tau\tau$ channel
in the MSSM search~\cite{ATLAS}. At high luminosity, pile-up effects
degrade the $\sla p_T$ resolution significantly, which results in a 
worse $\tau\tau$ invariant mass resolution. At a less significant level, 
a higher $p_T$ threshold for the minijet veto technique will 
increase the QCD and $t\bar t$ backgrounds. The $\tau$-identification 
efficiency is similar at high and low luminosity. 
We expect that the reduced performance at high 
luminosity can be compensated for by considering the additional channels 
$H\to\tau\tau \to e^+e^-\sla p_T,\;\mu^+\mu^-\sla p_T$. $Z+$jets and
$ZZ+$jets backgrounds (with $ZZ\to \ell^+\ell^-\nu\bar\nu$) are strongly 
suppressed by rejecting same flavor lepton pairs which are compatible 
with $Z$ decays ($m_{\ell\ell}= m_Z\pm 6$~GeV). Drell-Yan plus jets 
backgrounds are further reduced by requiring significant $\sla p_T$.
Since these analyses have not yet been performed, we 
use the predicted cross sections for only those two channels which have 
already been discussed in the literature 
and scale event rates to a combined 200~${\rm fb}^{-1}$ of data. 
Results are given in Table~\ref{table:WBF.tautau}.

\begin{table*}[t]
\vspace{0.2in}
\caption{Number of events expected for 
$qq\to qqH,\;H\to WW^{(*)}\to\mu^\pm e^\mp\sla p_T$ 
in 200~fb$^{-1}$ of data, and corresponding 
backgrounds~\protect\cite{RZ_WW}.
The expected relative statistical error on the signal cross section 
is given in the last line.}
\vspace{0.15in}
\label{table:WBF.WW}
\begin{tabular}{c|cccccccc}
$m_H$ &  120  &  130  &  140  &  150  &  160  &  170  &  180  &  190 \\
\hline
$N_S$ &  136  &  332  &  592  &  908  & 1460  & 1436  & 1172  &  832 \\
$N_B$ &  136  &  160  &  188  &  216  &  240  &  288  &  300  &  324 \\
$\Delta\sigma_H/\sigma_H$ &
       12.1\% & 6.7\% & 4.7\% & 3.7\% & 2.8\% & 2.9\% & 3.3\% & 4.1\%  \\
\end{tabular}
\end{table*}

The previous two weak boson channels allow reconstruction of the Higgs
resonance as an invariant mass peak. This is not the case for 
$H\to WW\to \ell^+\nu\ell^-\bar\nu$ as discussed previously for the 
inclusive search. The weak boson fusion 
channel can be isolated separately by employing forward jet tagging and
color singlet exchange isolation techniques in addition to tools like
charged lepton angular correlations which are used for the inclusive
channel. The corresponding parton level analysis for $qq\to qqH$, 
$H\to WW^{(*)}\to\mu^\pm e^\mp\sla p_T$ has been performed in
Ref.~\cite{RZ_WW} and we here scale the results to a total integrated 
luminosity of 200~${\rm fb}^{-1}$, which takes into account the availability 
of two detectors. As for the tau case, the analysis was done for low 
luminosity running conditions and somewhat higher backgrounds are expected
at high luminosity. On the other hand the $WW^{(*)}\to\mu^+\mu^-\sla p_T$ 
and $WW^{(*)}\to e^+e^-\sla p_T$ modes should roughly double the 
available statistics since very few signal events have lepton pair 
invariant masses compatible with $Z\to\ell\ell$ decays. Therefore our 
estimates are actually conservative. Note that the expected background 
for this weak boson fusion process is much smaller than for the 
corresponding
inclusive measurement. As a result modest systematic uncertainties will
not degrade the accuracy with which $\sigma(qq\to qqH)\cdot B(H\to 
WW^{(*)})$
can be measured. A 10\% systematic error on the background, double the error
assumed in the inclusive case, would degrade the statistical accuracy by, 
typically, a factor 1.2 or less. As a result, we expect that a very precise
measurement of  $\sigma(qq\to qqH)\cdot B(H\to WW^{(*)})$ can be performed 
at the LHC, with a statistical accuracy of order 5\% or even better in the 
mass range $m_H\geq 140$~GeV. Even for $m_H$ as low as 120~GeV a 12\% 
measurement is expected.

\section{Measurement of Higgs properties}
\label{sec3}

One would like to translate the cross section measurements of the various
Higgs production and decay channels into measurements of Higgs boson 
properties, in particular into measurements of the various Higgs boson
couplings to gauge fields and fermions. This translation requires knowledge
of NLO QCD corrections to production cross sections, information on the
total Higgs decay width and a combination of the measurements discussed
previously. The task here is to find a strategy for combining the 
anticipated
LHC data without undue loss of precision due to theoretical uncertainties
and systematic errors.

For our further discussion it is convenient to rewrite all Higgs boson 
couplings in terms of partial widths of various Higgs boson decay channels. 
The Higgs-fermion couplings $g_{Hff}$, for example, which in the SM are 
given
by the fermion masses, $g_{Hff} = m_f(m_H)/v$, can be traded for the
$H\to \bar ff$ partial widths,
\bq
\Gamma_f = \Gamma(H\to \bar ff) = c_f {g_{Hff}^2\over 8\pi}
\biggl( 1-{4m_f^2\over m_H^2} \biggr)^{3\over 2}\;m_H\;.
\eq
Here $c_f$ is the color factor (1 for leptons, 3 for quarks). Similarly the 
square of the $HWW$ coupling ($g_{HWW}=gm_W$ in the SM) or the $HZZ$ 
coupling is proportional to the partial widths $\Gamma_W=\Gamma(H\to WW^*)$ 
or $\Gamma_Z=\Gamma(H\to ZZ^*)$~\cite{Keung:1984hn}. Analogously we trade 
the squares of the effective $H\gamma\gamma$ and $Hgg$ couplings for 
$\Gamma_\gamma=\Gamma(H\to\gamma\gamma)$ and $\Gamma_g=\Gamma(H\to gg)$.
Note that the $Hgg$ coupling is essentially proportional to $g_{Htt}$, the
Higgs' coupling to the top quark.
 
The Higgs production cross sections are governed by the same squares of
couplings. This allows to write e.g. the $gg\to H$ production cross section
as~\cite{Barger:1987nn}
\bq
\sigma(gg\to H) = \Gamma(H\to gg){\pi^2\over 8m_H^3}\tau
\int_{\tau}^1 {dx\over x}g(x,m_H^2)g({\tau\over x},m_H^2)\;,
\eq
where $\tau = m_H^2/s$. Similarly the $qq\to qqH$ cross sections 
via $WW$ and $ZZ$ fusion are proportional to $\Gamma(H\to WW^*)$ and 
$\Gamma(H\to ZZ^*)$, respectively. In the narrow width approximation,
which is appropriate for the intermediate Higgs mass range considered here,
these production cross sections need to be multiplied by the branching
fractions for final state $j$, $B(H\to j)=\Gamma_j/\Gamma$, where $\Gamma$
denotes the total Higgs width. This means that the various cross section 
measurements discussed in the previous Section provide measurements of
various combinations $\Gamma_i\Gamma_j/\Gamma$. 

The production cross sections are subject to QCD corrections, which 
introduces theoretical uncertainties. While the $K$-factor for the gluon 
fusion process is large~\cite{HggNLO}, which suggests a sizable theoretical 
uncertainty on the production cross section, the NLO corrections to the weak 
boson fusion cross section are essentially identical to the ones encountered 
in deep inelastic scattering and are quite small~\cite{wbfNLO}. Thus we can 
assign a small theoretical uncertainty to the latter, of order 5\%, while 
we shall use a larger theoretical error for the gluon fusion process, of 
order 20\%~\cite{HggNLO}. 
The problem for weak boson fusion is that it consists of a mixture of
$ZZ\to H$ and $WW\to H$ events, and we cannot distinguish between the two 
experimentally. In a large class of models the ratio of $HWW$ and $HZZ$ 
couplings is identical to the one in the SM, however, and this includes the
MSSM. We therefore make the following $W,Z$-universality assumption:
\begin{itemize}
\item{}
The $H\to ZZ^*$ and $H\to WW^*$ partial widths are related by SU(2) as
in the SM, i.e. their ratio, $z$, is given by the SM value,
\bq
\Gamma_Z = z\; \Gamma_W = z_{SM}\;\Gamma_W \;.
\eq
\end{itemize}
Note that this assumption can be tested, at the 15-20\% level for 
$m_H>130$~GeV, by forming the ratio 
$B\sigma(gg\to H\to ZZ^*)/B\sigma(gg\to H\to WW^*)$, in which QCD 
uncertainties cancel (see Table~\ref{table:XYratio}).

With  $W,Z$-universality, the three weak boson fusion cross sections give us
direct measurements of three combinations of (partial) widths,
\ba
X_\gamma = {\Gamma_W\Gamma_\gamma\over \Gamma}\qquad &{\rm from}&\;\;
qq\to qqH,\; H\to\gamma\gamma \;, \\
X_\tau = {\Gamma_W\Gamma_\tau\over \Gamma}\qquad &{\rm from}&\;\;
qq\to qqH,\; H\to\tau\tau \;, \\
X_W = {\Gamma_W^2\over \Gamma}\quad\qquad &{\rm from}&\;\;
qq\to qqH,\; H\to WW^{(*)} \;,
\ea
with common theoretical systematic errors of 5\%. In addition the three gluon 
fusion channels provide measurements of 
\ba
Y_\gamma = {\Gamma_g\Gamma_\gamma\over \Gamma}\qquad &{\rm from}&\;\;
gg\to H\to\gamma\gamma \;, \\
Y_Z = {\Gamma_g\Gamma_Z\over \Gamma}\qquad &{\rm from}&\;\;
gg\to H\to ZZ^{(*)} \;, \\
Y_W = {\Gamma_g\Gamma_W\over \Gamma}\qquad &{\rm from}&\;\;
gg\to H\to WW^{(*)} \;,
\ea
with common theoretical systematic errors of 20\%. 

\begin{table}[th]
\caption{Summary of the accuracy with which various ratios of 
partial widths can be determined with 200 fb$^{-1}$ of data. The first two 
columns give the ratio considered and indicate the method by which it is 
measured. $Y_Z/Y_W$, for example, indicates a measurement of 
$\sigma B(H\to ZZ^*)/\sigma B(H\to WW^*)$ in gluon fusion, while $X_i$ 
ratios correspond to weak boson fusion (see text for details). The 
statistical combination of several channels for a given width ratio is 
indicated by $\oplus$. 5\% and 20\% theoretical uncertainties for weak 
boson and gluon fusion cross sections affect the mixed gluon/weak boson 
fusion ratios only, which are needed for a measurement of 
$\Gamma_g/\Gamma_W$. The effect of this systematic error is indicated 
in the last line.
}
\vspace{0.15in}
\label{table:XYratio}
\begin{tabular}{c|l|ccccccccc}
$m_H$ & & 100  &  110  &  120 &  130 &  140 &  150 &  160 &  170 &  180  \\
\hline
$z=\Gamma_Z/\Gamma_W$ & $Y_Z/ Y_W$ &
               &       & 48\% & 29\% & 19\% & 17\% & 15\% & 20\% & 17\%   \\
 & ${Y_Z\over Y_\gamma}{X_\gamma\over X_W}$ &
               &       & 30\% & 21\% & 19\% & 23\% &      &      &       \\
 & ${Y_Z\over Y_W}\oplus {Y_Z\over Y_\gamma}{X_\gamma\over X_W}$ &
               &       & 29\% & 19\% & 15\% & 14\% & 15\% & 20\% & 17\%   \\
\hline
$\Gamma_\gamma/\Gamma_W$ & ${Y_\gamma\over Y_W}\oplus{X_\gamma\over X_W}$ &
               &       & 16\% & 12\% & 11\% & 13\% &      &      &       \\
$\Gamma_\tau/\Gamma_W$ & ${X_\tau\over X_W}$ &
               &       & 15\% & 12\% & 14\% & 21\% &      &      &       \\
$\Gamma_\tau/\Gamma_\gamma$ & ${X_\tau\over X_\gamma}$ &
          20\% &  16\% & 15\% & 16\% & 18\% & 27\% &      &      &       \\
\hline
$\Gamma_g/\Gamma_W$ & ${Y_\gamma\over X_\gamma}\oplus{Y_W\over X_W}$ &
          22\% &  18\% & 15\% & 13\% & 12\% & 13\% &  8\% &   9\% & 14\%  \\
       & ${Y_\gamma\over X_\gamma}\oplus{Y_W\over X_W}\oplus 21\%$ &
          30\% &  27\% & 25\% & 24\% & 24\% & 24\% & 22\% & 22\% & 25\%  \\
\end{tabular}
\end{table}

The first precision test of the Higgs sector is provided by taking ratios 
of the $X_i$'s and ratios of the $Y_i$'s. In these ratios the QCD 
uncertainties, and all other uncertainties related to the initial state, 
like luminosity and pdf errors, cancel.  Beyond testing $W,Z$-universality,
these ratios provide useful 
information for Higgs masses between 100 and 150~GeV and 120 to 150~GeV, 
respectively, where more than one channel can be observed in the weak 
boson fusion and gluon fusion groups. Typical errors on these cross 
section ratios are expected to be in the 15 to 20\% range (see 
Table~\ref{table:XYratio}).
Accepting an additional systematic error of about 20\%, a measurement 
of the ratio $\Gamma_g/\Gamma_W$, which determines the $Htt$ to $HWW$ 
coupling ratio, can be performed, by measuring the cross section 
ratios $B\sigma(gg\to H\to\gamma\gamma)/\sigma(qq\to 
qqH)B(H\to\gamma\gamma)$
and  $B\sigma(gg\to H\to WW^*)/\sigma(qq\to qqH)B(H\to WW^*)$. Expected
accuracies are listed in Table~\ref{table:XYratio}. In these estimates 
the systematics coming from understanding detector acceptance is not 
included. 

Beyond the measurement of coupling ratios, minimal additional assumptions 
allow an indirect measurement of the total Higgs width. First of all, the 
$\tau$ partial width, properly normalized, is measurable with an accuracy of 
order 10\%. The $\tau$ is a third generation fermion with isospin 
$-{1\over 2}$, just like the $b$-quark. In all extensions of the SM with 
a common source of lepton and quark masses, even if generational symmetry 
is broken, the ratio of $b$ to $\tau$ Yukawa couplings is given by 
the fermion mass ratio. We thus assume, in addition to $W,Z$-universality,
that
\begin{itemize}
\item{}
The ratio of $b$ to $\tau$ couplings of the Higgs is given by their mass 
ratio, i.e.
\bq
y = {\Gamma_b\over \Gamma_\tau} = 3c_{QCD}{g_{Hbb}^2\over g_{H\tau\tau}^2}
= 3c_{QCD}{m_b^2(m_H)\over m_\tau^2}\;,
\eq
where $c_{QCD}$ is the known QCD and phase space correction factor.

\item{}
The total Higgs width is dominated by decays to $\bar bb$, $\tau\tau$, 
$WW$, $ZZ$, $gg$ and $\gamma\gamma$, i.e. the branching ratio for 
unexpected channels is small:
\ba
\epsilon = 1-\biggl(&& B(H\to b\bar b)+B(H\to \tau\tau)+B(H\to WW^{(*)})+
\nonumber \\
&& B(H\to ZZ^{(*)})+B(H\to gg)+B(H\to \gamma\gamma)\biggr) \qquad \ll 1\;.
\ea
\end{itemize}
Note that, in the Higgs mass range of interest, these two assumptions are 
satisfied for both CP even Higgs bosons in most of the MSSM parameter space.
The first assumption holds in the MSSM at tree level, but
can be violated by large squark loop contributions, in particular for small
$m_A$ and large $\tan\beta$~\cite{carena,MSSMcalc}. 
The second assumption might be 
violated, for example, if the $H\to\bar cc$ partial width is exceptionally 
large. However, a large up-type Yukawa coupling would be noticeable in 
the $\Gamma_g/\Gamma_W$ coupling ratio, which measures the $Htt$ coupling.

\begin{figure}[thb]
\begin{center}
\includegraphics[width=9.0cm,angle=0]{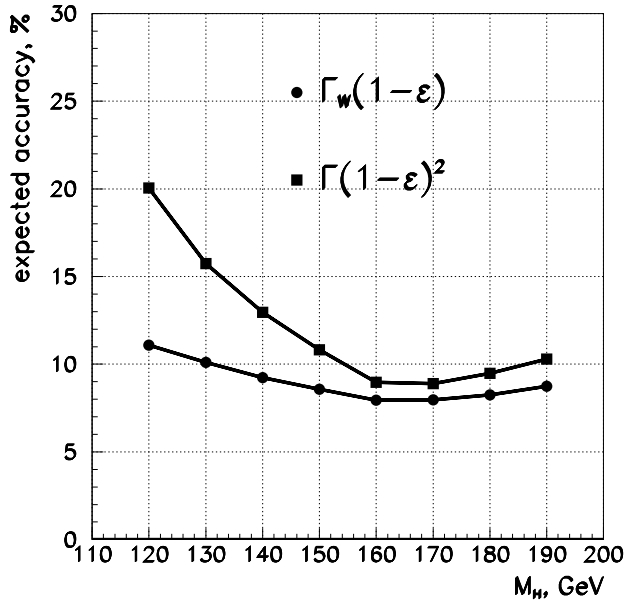}
\end{center}
\vspace*{0.2cm}
\caption{Expected accuracy with which the Higgs boson width can be measured 
at the LHC, with 100~fb$^{-1}$ of data in each experiment. Results are shown
for the extraction of the the $H\to WW$ partial with, $\Gamma_W$, and
and the total Higgs boson width, $\Gamma$. $\epsilon$ is the sum of the
residual (small) branching ratios of unobserved channels, mainly 
$H\to c\bar c$ (see text for detail).}
\label{fig:Widthaccurcy}
\vspace*{0.2cm}
\end{figure}

With these assumptions consider the observable
\begin{eqnarray}
\tilde\Gamma_W &=& X_\tau(1+y) + X_W(1+z) + X_\gamma + \tilde X_g 
\nonumber \\
&=& \biggl(\Gamma_\tau+\Gamma_b +\Gamma_W +\Gamma_Z+
\Gamma_\gamma+\Gamma_g\biggr){\Gamma_W\over\Gamma}
=(1-\epsilon)\Gamma_W  \;,
\end{eqnarray}
where $\tilde X_g = \Gamma_g\Gamma_W/\Gamma$ is determined by combining 
$Y_W$
and the product $Y_\gamma X_W/X_\gamma$. $\tilde\Gamma_W$ provides a lower 
bound on $\Gamma(H\to  WW^{(*)})=\Gamma_W$. Provided $\epsilon$ is small 
($\epsilon < 0.1$ suffices for practical purposes), the determination 
of $\tilde\Gamma_W$ provides a direct measurement of the $H\to WW^{(*)}$ 
partial width. Once $\Gamma_W$ has been determined, the total width of the 
Higgs boson is given by
\bq
\label{eq:Gamma_tot}
\Gamma = {\Gamma_W^2\over X_W}={1\over X_W}
\biggl(X_\tau(1+y) + X_W(1+z) + X_\gamma + \tilde X_g \biggr)^2
{1\over (1-\epsilon)^2}\; .
\eq
For a SM-like Higgs boson the Higgs width is dominated by the $H\to\bar bb$
and $H\to WW^{(*)}$ channels. Thus, the error on $\tilde\Gamma_W$ is
dominated by the uncertainties of the $X_W$ and $X_\tau$ measurements 
and by the theoretical uncertainty on the $b$-quark mass, which enters 
the determination of $y$ quadratically. According to the Particle Data Group,
the present 
uncertainty  on the $b$ quark mass is about $\pm 3.5\%$~\cite{pdg98}. 
Assuming a luminosity error of $\pm 5\%$ in addition to the theoretical 
uncertainty of the weak boson fusion cross section of $\pm 5\%$, the 
statistical
errors of the $qq\to qqH,\;H\to \tau\tau$ and $qq\to qqH, H\to WW$
cross sections of Tables~\ref{table:WBF.tautau} and \ref{table:WBF.WW} 
lead to an expected accuracy of the $\tilde\Gamma_W$ determination 
of order 10\%. 
More precise estimates, as a function of the Higgs boson mass, are shown
in Fig.~\ref{fig:Widthaccurcy}.

The extraction of the total Higgs width, via Eq.~(\ref{eq:Gamma_tot}),
requires a measurement of the $qq\to qqH, H\to WW^{(*)}$ cross section,
which is expected to be available for $m_H\gsim 115$~GeV~\cite{RZ_WW}.
Consequently, errors are large for Higgs masses close to this lower limit
(we expect a relative error of $\approx 20\%$ for $m_H=120$~GeV and 
$\epsilon<0.05$). But for Higgs boson masses around the $WW$ threshold, 
$\Gamma(1-\epsilon)^2$ can be determined with an error of about 10\%. 
Results are shown in Fig.~\ref{fig:Widthaccurcy} and look highly promising.

\section{Summary}
\label{sec4}

In the last section we have found that various ratios of Higgs partial 
widths
can be measured with accuracies of order 10 to 20\%, with an integrated 
luminosity of 100 fb$^{-1}$ per experiment. This translates into 5 to 10\%
measurements of various ratios of coupling constants. The 
ratio $\Gamma_\tau/\Gamma_W$ measures the coupling of down-type fermions
relative to the Higgs couplings to gauge bosons. To the extent that the
$H\gamma\gamma$ triangle diagrams are dominated by the $W$ loop,
the width ratio $\Gamma_\tau/\Gamma_\gamma$ measures the same relationship.
The fermion triangles leading to an effective $Hgg$ coupling are expected 
to be dominated by the top-quark, thus, $\Gamma_g/\Gamma_W$ probes the 
coupling of up-type fermions relative to the $HWW$ coupling.
Finally, for Higgs boson masses above $\approx 120$~GeV,
the absolute normalization of the $HWW$ coupling is accessible
via the extraction of the $H\to WW^{(*)}$ partial width
in weak boson fusion.

Note that these measurements test the crucial aspects of the Higgs sector.
The $HWW$ coupling, being linear in the Higgs field, 
identifies the observed Higgs boson as the scalar 
responsible for the spontaneous breaking of $SU(2)\times U(1)$: a 
scalar without a vacuum expectation value couples to gauge bosons only 
via $HHWW$ or $HHW$ vertices at tree level, i.e. the interaction is 
quadratic in scalar fields.
The absolute value of the $HWW$ coupling, as compared to the SM expectation,
reveals whether $H$ may be the only mediator of spontaneous symmetry breaking
or whether additional Higgs bosons await discovery. Within the framework of 
the MSSM this is a measurement of $|\sin (\beta-\alpha)|$, at the 
$\pm 0.05$ level. 
The measurement of the ratios of $g_{Htt}/g_{HWW}$ and 
$g_{H\tau\tau}/g_{HWW}$ then probes the mass generation of both up and down
type fermions. 

The results presented here constitute a first look only at the issue of
coupling extractions for the Higgs. This is the case for the weak boson fusion
processes in particular, which prove to be extremely valuable if not 
essential. Our analysis
is mostly an estimate of statistical errors, with some rough estimates
of the systematic errors which are to be expected for the various 
measurements of (partial) widths and their ratios. A number of issues need 
to be addressed in further studies, in particular with regard to the weak
boson fusion channels.

\begin{itemize}
\item[(a)]
The weak boson fusion channels and their backgrounds have only been studied 
at the parton level, to date. Full detector level simulations, and 
optimization of strategies with more complete detector information is 
crucial for further progress. 

\item[(b)] 
A central jet veto has been suggested as a powerful tool to suppress 
QCD backgrounds to the color singlet exchange processes which we call 
weak boson fusion. The feasibility of this tool and its reach need to be 
investigated in full detector studies, at both low and high luminosity.

\item[(c)]
In the weak boson fusion studies of $H\to WW$ and $H\to\tau\tau$ decays,
double leptonic $e^+e^-\sla p_T$ and $\mu^+\mu^-\sla p_T$ signatures have 
not yet been considered. Their inclusion promises to almost double the 
statistics available for the Higgs coupling measurements, at the price
of additional $ZZ+$jets and Drell-Yan plus jets backgrounds which are
expected to be manageable.

\item[(d)]
Other channels, like $WH$ or $t\bar tH$ associated production with 
subsequent
decay $H\to \bar bb$ or $H\to\gamma\gamma$, provide additional information
on Higgs coupling ratios, which complement our analysis at small Higgs mass
values, $m_H\lsim 120$~GeV~\cite{CMS,snowmass_H}. These channels need to be 
included in the analysis.

\item[(e)]
Much additional work is needed on more reliable background determinations. 
For the $H\to WW^{(*)}\to \ell^+\ell^{'-}\sla p_T$ channel in particular, 
where no narrow Higgs resonance peak can be reconstructed, a precise
background estimate is crucial for the measurement of Higgs couplings. 
Needed
improvements include NLO QCD corrections, single top quark production 
backgrounds, the combination of shower Monte Carlo programs with higher
order QCD matrix element calculations and more.

\item[(f)]
Both in the inclusive and WBF analyses any given channel contains a 
mixture of events from $gg\to H$ and $qq\to qqH$ production processes.
The determination of this mixture adds another source of systematic 
uncertainty, which was not included in the present study.
In ratios of $X$ observables (or of different $Y_i$) these uncertainties
largely cancel, except for the effects of acceptance variations due to
different signal selections. Since an admixture from the wrong production 
channel is expected at the 10 to 20\% level only, these systematic errors 
are not expected to be serious. 

\item[(g)]
We have only analyzed the case of a single neutral, CP even Higgs 
resonance with couplings which are close to the ones predicted in the SM.
While this case has many applications, e.g. for the large $m_A$ region of 
the MSSM, more general analyses, in particular of the MSSM case, are 
warranted and highly promising. 

\end{itemize}

While much additional work is needed, our study clearly shows that the 
LHC has excellent potential to provide detailed and accurate information 
on Higgs boson interactions. The observability of the Higgs boson at the LHC
has been clearly established, within the SM and extensions like the MSSM.
The task now is to sharpen the tools for accurate measurements of 
Higgs boson properties at the LHC.

\acknowledgements
We would like to thank the organizers of the Les Houches Workshop, where
this work was initiated, for getting us together in an inspiring atmosphere.
Useful discussions with M.~Carena, A.~Djouadi, K.~Jakobs and G.~Weiglein
are gratefully acknowledged.
We thank CERN for the hospitality extended to all of us during various 
periods of this work. 
The research of E.~R.-W. was partially supported by the Polish Government 
grant KBN 2P03B14715, and by the Polish-American Maria Sk\c lodowska-Curie 
Joint Fund II in cooperation with PAA and DOE under project PAA/DOE-97-316. 
The work of D.~Z. was supported in part by the University of Wisconsin Research
Committee with funds granted by the Wisconsin Alumni Research Foundation and
in part by the U.~S.~Department of Energy under Contract
No.~DE-FG02-95ER40896.


\bibliographystyle{plain}

\end{document}